\documentclass[preprint]{aastex}	 
\usepackage{epsfig,emulateapj5}

\newcommand{\km}{${\rm km\,s}^{-1}$}
\newcommand{\fuse}{\em FUSE\/}
\newcommand\sk[2]{Sk~{$#1^{\circ}#2$}}
\newcommand{\vinf}{${v}_\infty$}

\slugcomment{Version of \today}
\shorttitle{\ion{O}{6} wind variability in hot stars and impact on interstellar \ion{O}{6} 
            measurements}
\shortauthors{Lehner et al.}
\begin{document}

\title{ The Influence of Stellar Wind Variability on Measurements
        of Interstellar  {\ion{O}{6}} Along Sightlines to Early-Type Stars}

\author{N.\ Lehner,\altaffilmark{1}
        A.\ W.\ Fullerton,\altaffilmark{2,1}
        K.\ R.\ Sembach,\altaffilmark{1}
        D.\ L.\ Massa,\altaffilmark{3}
	E.\ B.\ Jenkins\altaffilmark{4}}
   
\altaffiltext{1}{Department of Physics and Astronomy, 
                 The Johns Hopkins University,
                 3400 N. Charles Street, 
		 Baltimore, MD 21218.
		 {Email: \rm{nl@pha.jhu.edu.}}  }

\altaffiltext{2}{Department of Physics and Astronomy, 
                 University of Victoria, 
		 P.O. Box 3055,
                 Victoria, BC V8W 3P6, 
		 Canada.}

\altaffiltext{3}{Emergent IT,
                 NASA's Goddard Space Flight Center, 
		 Code 681, Greenbelt, MD 20771.}

\altaffiltext{4}{Princeton University Observatory,
                 Princeton, NJ 08544}

\begin{abstract}

A primary goal of the {\fuse} mission is to understand the origin 
of the {\ion{O}{6}} ion in the interstellar  medium of the Galaxy and 
the Magellanic Clouds.
Along sightlines to OB-type stars, these interstellar  components are
usually blended with {\ion{O}{6}} stellar wind profiles, which
frequently vary in shape.
In order to assess the effects of this time-dependent blending
on measurements of the interstellar  {\ion{O}{6}} lines, we have undertaken a
mini-survey of repeated observations toward OB-type stars in 
the Galaxy and the Large Magellanic Cloud.
These sparse time series, which consist of 2--3 observations separated 
by intervals ranging from a few days to several months, show that wind 
variability occurs commonly in {\ion{O}{6}} ($\sim$60$\%$ of a 
sample of 50 stars), as indeed it does in other 
resonance lines. However, in the interstellar {\ion{O}{6}} $\lambda$1032 region,
the {\ion{O}{6}} $\lambda$1038 wind varies only in $\sim$30$\%$ of the cases. 
By examining cases exhibiting large amplitude variations, we conclude that
stellar-wind variability {\em generally} introduces negligible uncertainty for
single interstellar  {\ion{O}{6}} components along Galactic lines of sight,
but can result in substantial errors in measurements of broader components
or blends of components like those typically observed toward stars in the Large Magellanic Cloud.
Due to possible contamination by discrete absorption components in the
stellar {\ion{O}{6}} line, stars with terminal velocities greater than
or equal to the doublet separation (1654 \km) should be treated with care.
\end{abstract}


\section{Introduction}

The {\ion{O}{6}} $\lambda\lambda$1032, 1038 resonance doublet
is an excellent diagnostic of high-energy processes ($\ge 113.9$ eV) in rarefied media
because of the large cosmic abundance of oxygen and the
substantial energy required to strip five electrons from O atoms.
A key program for the {\it Far Ultraviolet Spectroscopic Explorer\/} 
({\fuse}) mission is to measure the strength of this
doublet along many sightlines, in order to determine the
origin and evolution of hot gas in the interstellar  medium
\citep[ISM;~see,~e.g.,][]{savage00,sembach00}.

However, {\ion{O}{6}} also occurs in the stellar winds of the 
OB-type stars that frequently serve to illuminate the intervening ISM.
Its presence in these winds is surprising, since the photospheric radiation 
fields of only the hottest stars produce atmospheric {\ion{O}{6}}.
A current explanation is that winds contain ensembles of strong shocks,
which might be produced by the strong line-driven instability
\citep{owocki88}.
The X-rays created by these shocks produce {\ion{O}{6}} from
the dominant {\ion{O}{4}} ion via Auger ionization \citep{cassinelli79}.
Consequently, the shapes and strengths of the {\ion{O}{6}} wind
profiles are largely determined by the distribution, strength,
and time-dependent behavior of the shocks.
Indeed, previous work with {\it Copernicus} 
{\citep[see,~e.g.,][]{york77,snow80}} and early results from 
{\fuse} {\citep{massa00}} confirm that the {\ion{O}{6}} wind lines of 
early-type stars are variable on a variety of time scales, as are most 
other wind features; see \citet{prinja98a} for a recent review.
As a result, undulations in the broad P-Cygni wind profiles blend with the 
interstellar components in a complicated, time-dependent way, which may
compromise measurements of interstellar {\ion{O}{6}} by unknown amounts.

To assess the influence of stellar wind variability on measurements
of interstellar  {\ion{O}{6}}, we have used {\fuse} to obtain
2--3 observations for each object in a sample of $\sim$50 OB-type stars 
in the Galaxy and Large Magellanic Cloud (LMC).
These sparse time series exhibit variations over intervals ranging
from about a day to several months.
Preliminary results suggest that stellar wind variability is often found
in {\ion{O}{6}}, with no apparent differences between the Galaxy
and the LMC.
In this {\it Letter\/} we illustrate the influence of the variability
on interstellar  measurements along five sightlines, which typify the problems 
that may be encountered.

\section{Observations and Analysis}
The design and performance of the {\fuse} spectrograph have been
described by {\citet{moos00}} and {\citet{sahnow00}}, respectively.
Table~\ref{t1} summarizes the repeated {\fuse} observations for the
five sightlines discussed here.
The observations of HD\,97913 were obtained in histogram mode;
all others were obtained in time-tag mode.
All observations were obtained through the $30\arcsec \times 30\arcsec$ apertures.
Standard processing with the current version of the calibration
pipeline software ({\sc calfuse} v1.8.7) was used to extract
and calibrate the spectra. 
The extracted spectra associated with separate exposures of
a given observation were aligned by cross-correlating the
positions of strong interstellar lines (other than {\ion{O}{6}}).
Since these subexposures did not exhibit significant variations,
they were coadded and rebinned to a spectral resolution
of $\sim$15,000.

Fig.~\ref{fig1} shows the {\fuse} spectra of the LMC star
{\sk{-69}{124}} and summarizes the method used to detect stellar
wind variability in these sparse time series.
    (1) Since the flux is well calibrated,
       two spectra of the same
       object taken at different times were overplotted and directly compared
       (top panel).
   (2) To ensure that the wind variations are genuine, we require that they
     are similar in both transitions of the doublet (bottom panel) and
     in data from different detector segments.
Since the \ion{O}{6} $\lambda$1032 line is intrinsically
stronger than the $\lambda$1038 line, the amplitude of any fluctuation will
generally be larger in the blue component.
Fig.~\ref{fig1} shows clearly the dramatic effect that the stellar wind 
variability might have on interstellar  measurements. If the \ion{O}{6} 
$\lambda$1038 Galactic feature is detected, the LMC component is not; but 
even more striking is the absence of any obvious interstellar \ion{O}{6} $\lambda$1032 component.

To quantify the influence of the stellar wind variability on the interstellar measurements,
we consider three other stars that exhibit large variations in the vicinity of the
interstellar {\ion{O}{6}} $\lambda$1032 feature (Table~\ref{t1} and Fig.~\ref{fig2}).
Typically, the signal-to-noise ratios per resolution element lie between 
15 and 30 for these spectra.
We have fitted the apparent stellar continuum of each observation independently;
i.e., we did not fit the continua with a preconceived notion based on the other 
spectrum. For that reason, the chosen continua look different between panels in Fig.~\ref{fig2}.
The criterion to define the continuum was that the \ion{O}{6} $\lambda$1032
full width (FW) should not be larger than about $[-70,+70]$ \km\ for the Galactic
stars \citep{savage00}, while for the LMC it should be approximately in the range
of $[-80,+350]$ \km\ (Howk, priv. comm., 2001),
with small scatter from one sightline to another.
The \ion{O}{6} $\lambda$1032 line is also bracketed by two H$_2$ lines, 
$(6-0)$ $P(3)$  $\lambda$1031.19 and $R(4)$ $\lambda$1032.35, which often complicate 
continuum placement. 
Although a detailed H$_2$ model based on all the transitions in the {\fuse} bandpass
could in principle constrain the influence of these lines 
on the local continuum, for this study we have simply interpolated over them.
Moreover, if the interstellar \ion{O}{6} $\lambda$1038 feature is resolved,
it should always be compared to \ion{O}{6} $\lambda$1032, though generally 
\ion{O}{6} $\lambda$1038 is very difficult to discern due to blending with
\ion{C}{2}, \ion{C}{2*}, and H$_2$; this is the case for all the stars discussed here.
The continuum was fitted by a Legendre polynomial of degree
$d$ (see Table~\ref{t1}). 
Fig.~\ref{fig2} shows very complex profiles and therefore the order of 
the polynomial to fit the continuum needed to be relatively large. 
Once the continuum was considered satisfactory, the equivalent width
and column density were calculated by directly integrating the 
\ion{O}{6} $\lambda$1032 intensity and apparent column density, 
$ N_a(v) \equiv 3.768 \times 10^{14} \tau_a(v)/[f \lambda_0(\rm\AA)]$
\citep{savage91} where $f=0.132$ and $\lambda_0 = 1031.926$. 
The results are given in Table~\ref{t1}. 
The statistical errors and errors on continuum placement, 
presented in Table~\ref{t1}, were obtained following the method outlined by \citet{sembach92}. 
The continuum placement errors are comparable to or larger than the
statistical errors and do not reflect systematic differences that
result from different choices for the wavelength intervals defining
the (unknown) location of the true continuum.

For the two Galactic stars, there is good agreement between the
absolute measurements of the interstellar {\ion{O}{6}} features, though the 1$\sigma$ errors can be 
quite large (see Table~\ref{t1}). 
However, even though the H$_2$ $P(3)$ and {\ion{O}{6}} features are similar for 
both the normalizations illustrated in Fig.~\ref{fig2}, the depth of the 
H$_2$ $R(4)$ line is quite different, which suggests that the continuum is not 
well defined on the red side.
The discrepancy is particularly obvious in the case of HD\,168941.
For HD\,97913, the two normalized profiles have noticeable differences, even
though the measured equivalent widths agree within the uncertainties.
For this star, we did not account properly for the \ion{Cl}{1} 
feature; this also explains why the H$_2$ $P(3)$ line appears to be overnormalized.
The situation was even worse for the second observation and therefore 
we concentrated on a smaller part of the spectrum.
The sightline toward the LMC star {\sk{-67}{191}} shows large discrepancies
depending on which normalization is adopted, so that the resultant equivalent
widths do not agree within the estimated $1\sigma$ uncertainties.

\section{Discussion}
\begin{figure*}[!th]
\begin{center}
\includegraphics[width=8.8 truecm]{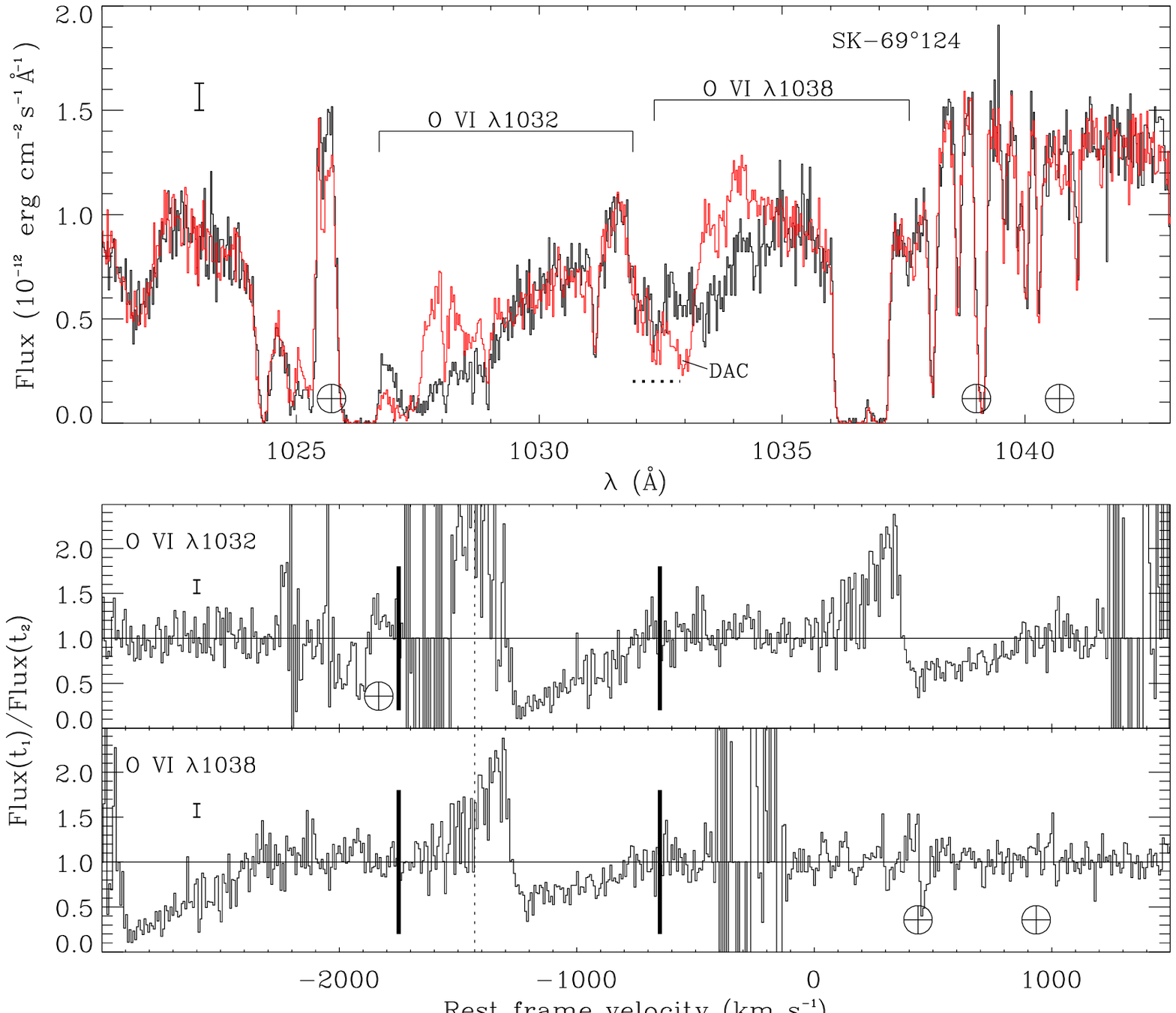}
\caption{} 
\label{fig1}
\end{center}
\end{figure*}

Stellar wind variability can have serious consequences for
interstellar \ion{O}{6} studies:
(1) In some instances, it can  make the interstellar \ion{O}{6} identification very difficult
    (see Fig.~\ref{fig1}).
(2) It complicates significantly the stellar continuum placement
which increases the uncertainty in determining the interstellar line strength, or
even gives erroneous results in situations where 
multiple interstellar components produce broad features.
Since the interstellar {\ion{O}{6}} FW toward LMC stars is typically about three times 
that observed in Galactic stars, analysis of LMC sightlines are particularly 
susceptible to these uncertainties.  
While this study shows that stellar wind variability is {\em generally}
negligible for studies of interstellar O VI along sightlines toward
Galactic OB-type stars, systematic errors in continuum placement
can remain large for complicated continua.
Modeling of stellar wind profiles could in principle 
help to constrain the shape of 
the local continuum, but at present the fits are uncertain for Galactic stars 
(due mainly to the H$_2$ contamination), and moreover, do not yet incorporate 
time-dependent behavior in a self-consistent way. 
However, since these cases show wind variations very close to
the interstellar \ion{O}{6} $\lambda$1032 line, they should be considered
as ``worst case'' scenarios.
Preliminary results from our mini-survey for {\ion{O}{6}} variability
(Lehner et al. 2001, in prep.) indicate that at least $\sim$60$\%$ of our sample
exhibits {\ion{O}{6}} wind variability; {\em but} less than $\sim$30$\%$ (14/50) of
the total sample exhibits variations in the region of the interstellar features, i.e.,
$-600 < v < 600$ \km, at the time of the survey. 
The relevant velocity ranges are defined by the separations
between the \ion{O}{6} doublet and \ion{O}{6} $\lambda$1032 and \ion{H}{1} 
Ly $\beta$, 1654 and 1805 \km, respectively.
For 10 of these 14 stars, the terminal velocity ({\vinf}) is known:
7/10 have roughly $ 1665 \lesssim v_\infty \lesssim 2135$ \km,
while 3/10 have lower {\vinf} and can blend with high velocity interstellar components. 

\begin{figure*}[!th]
\begin{center}
\includegraphics[width=18.5 truecm]{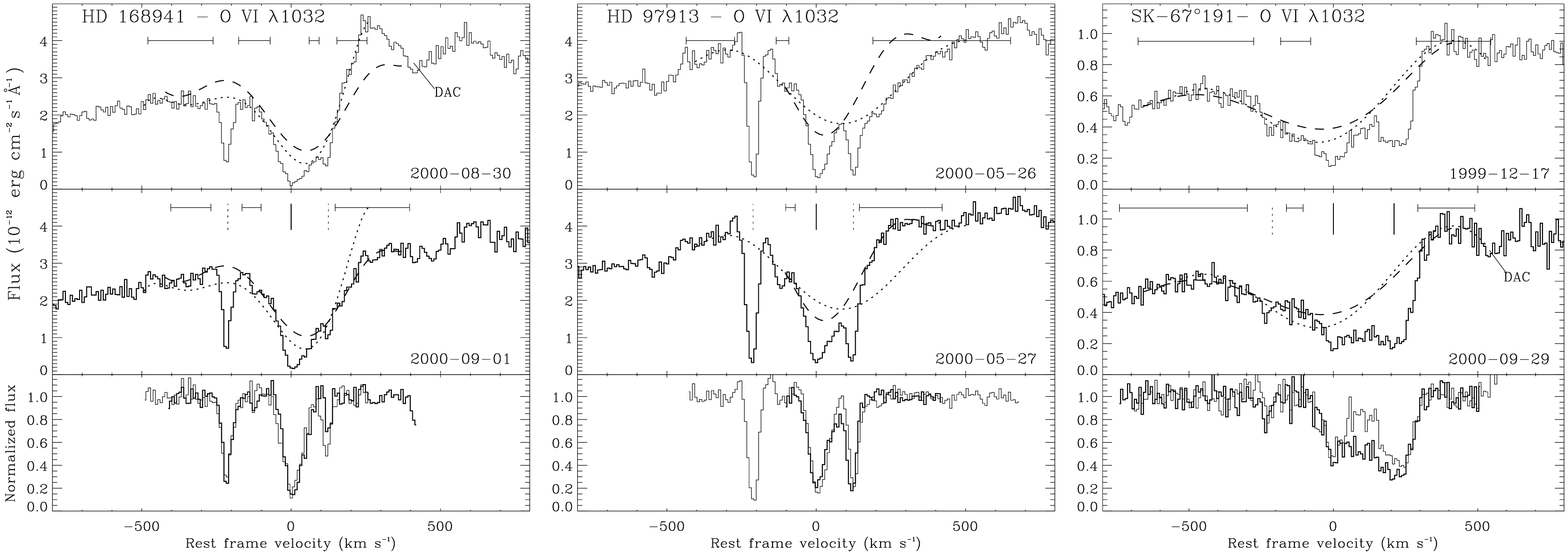}
\caption{
  Top and Middle:  Continuum placement in the region of
  \ion{O}{6} $\lambda$1032. 
  Horizontal bars show the regions used to define the continuum.
  The dotted line corresponds to the continuum of the spectrum in the
  top panel, while the dashed line is the continuum of the spectrum in
  the middle panel.
  Bottom: Normalized fluxes resulting from the different continuum fits.
  The thicker line corresponds to the continuum from the middle panel.
  Dotted tick marks indicate the positions of H$_2$ lines;
  solid tick marks show the positions of the {\ion{O}{6}} Galactic and
  LMC components ($\sim$ 0 and 200 \km, respectively).
  The feature at $\sim-$135~\km\ in the spectrum of HD\,97913 is
  {\ion{Cl}{1}} $\lambda$1031.3} 
\label{fig2}
\end{center}
\end{figure*}

\begin{figure*}[!th]
\begin{center}
\includegraphics[width=8.8 truecm]{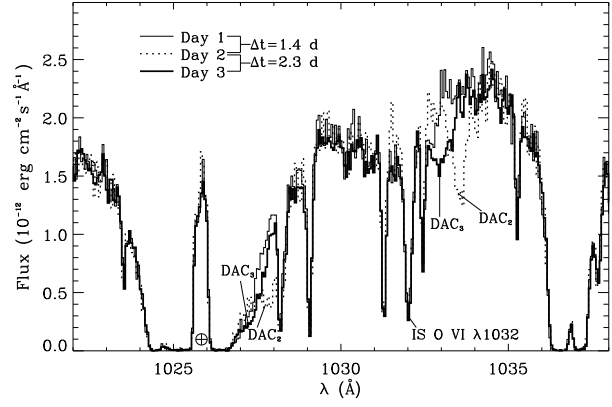}
\caption{Spectra of HD\,91597 showing the propagation of a DAC in the 
         {\ion{O}{6}} wind profile of $\lambda1032$ ($\lambda \lesssim 1028$ \AA)
	 and $\lambda1038$ ($\lambda \lesssim 1034$ \AA), corresponding to velocities
	 of $\sim$$-1170$ \km\ (day 2) and $\sim$$-1370$ \km\ (day 3). The 
	 DAC accelerates slowly ($\sim$10$^{-3}$ \km), which is typical
	 of the motion of DACs observed in the winds of other stars (Prinja 1998).} 
\label{fig3}
\end{center}
\end{figure*}

Although the nature of the stellar wind variations cannot be determined
from the mini-survey, they are likely related to the discrete
absorption components (DACs) that are commonly observed in the wind lines
of hot stars \citep{prinja98a}.
DACs are particularly strong near {\vinf}. When $v_\infty \ll 1650$ \km, the interstellar
measurement will not be affected by DACs
(provided that $v_{\rm ISM}$ is not much greater than 0 \km).
But when  $v_\infty \gtrapprox 1650$ \km, contamination by DACs is possible.
When $v_\infty \gg 1650$ \km, the effect of the DACs on the local continuum
is much less, since they change during their blueward motion from broad, weak 
features to a deep, sharp one near {\vinf} \citep{prinja98a}, assuming that 
\ion{O}{6} DACs behave as for the other ions.
It is also important to note that the separation
between \ion{O}{6} $\lambda$1032 and the edge of the strongly saturated part of \ion{H}{1}
Ly $\beta$ can be $\ll 1800$ \km, and therefore information
from the DACs in {\ion{O}{6}} $\lambda$1032 can be lost.
Fig.~\ref{fig3} shows the blueward motion of a DAC in the wind of the
Galactic star HD\,91597.
Three other sightlines presented here show an absorption
feature on the red side of \ion{O}{6} $\lambda$1032 (see Figs.~\ref{fig1} \& \ref{fig2}).
In the more extended sample that will appear in our later paper, 
only one case has a FW comparable to a Galactic interstellar \ion{O}{6}
line, while all others have typically $200 \lesssim {\rm FW} \lesssim 400$  \km.
Although we are not able to quantify further the effects of DACs from short time
series of many different objects, it is clear that their presence in
{\ion{O}{6}} wind profiles can complicate continuum placement
(see Fig.~\ref{fig2}) and in the worst case situations even produce erroneous measurements.

\acknowledgements
We thank Chris Howk and Dave Bowen for insightful comments on an early draft of this Letter,
and Don York for initially suggesting a study to gauge the
impact of stellar variability on measurements of interstellar
\ion{O}{6} absorption.
This work is based on data obtained for the Guaranteed Time Team by the 
NASA-CNES-CSA FUSE mission operated by the Johns Hopkins University. 
Financial support to U. S. participants has been provided by NASA 
contract NAS5-32985.

\newpage

\begin{deluxetable}{llcccccccc}
\tablecolumns{10}
\tablewidth{0pc} 
\tablecaption{Program star summary and interstellar {\ion{O}{6}} measurements \label{t1}}
\tablehead{\colhead{Star}    &   \colhead{Sp. Type}&   \colhead{\vinf} &\colhead{UT-Date}&\colhead{$\Delta t$$^a$}&\colhead{$N_{\rm exp}$}&\colhead{$t_{\rm exp}$} &\colhead{$d$$^b$} &\colhead{$W_\lambda$$^c$} &\colhead{$N_a$$^c$} 	\\ 
\colhead{}    &   \colhead{}&  \colhead{(\km)} & \colhead{}&  \colhead{(days)} & \colhead{}& \colhead{(s)} &\colhead{}&\colhead{(m\AA)}& \colhead{(dex)}}
\startdata
\sk{-67}{191}		&  O8.5\,V      & 1750$^d$	& 1999-12-17	 & 287.0& 2 & 8460 	&  6    & $440 \pm 23 \pm 53$  & $14.67 \pm 0.07 $ \\
			& 	        & 	 	& 2000-09-29	 &	& 1 & 6569 	&  5    & $609 \pm 23 \pm 45$  & $14.85 \pm 0.05 $ \\
\sk{-69}{124}		&  O9\,II       & 1430$^d$	& 2000-10-03	 & 62.1 & 1 & 2120$^e$ 	&       &	&	\\
			& 	        & 	 	& 2000-12-04	 &	& 2 & 4220$^e$ 	&       &	&	\\
HD\,168941		&  O9.5 II-III  & 1795$^f$	& 2000-08-30	 & 2.2  & 2 & 4110 	&  6    & $221 \pm 24 \pm 25$  & $14.46 \pm 0.11 $  \\
			&  	        & 	 	& 2000-09-01	 &	& 2 & 3982 	&  7    & $221 \pm 16 \pm 51$  & $14.49 \pm 0.13 $ \\
HD\,97913		&  B0.5\,IVn    & \nodata	& 2000-05-26	 & 1.1  & 8 & 5996 	&  6    & $195 \pm 9 \pm 13$   & $14.43 \pm 0.05 $ \\
			&	        & 	 	& 2000-05-27	 &	& 12& 9980     	&  3    & $204 \pm 11 \pm 40$  & $14.40 \pm 0.12 $ \\
HD\,91597		&  B1\,IIIne    & \nodata 	& 2000-02-04	 & 1.4  & 2 & 2030$^e$ 	&       &	&	\\
			&	        & 	 	& 2000-02-06	 & 2.3  & 2 & 2100$^e$ 	&       &	&	\\
			&	        & 	 	& 2000-02-08	 &	& 4 & 5629$^e$ 	&       &	&	\\
\enddata
\tablecomments{
($a$) $\Delta t$: time between successive observations.
($b$) Order of the polynomial used to fit the local continuum
adjacent to the interstellar \ion{O}{6} feature.
($c$) Errors are $1\sigma$. For the equivalent widths, the first error is statistical, while 
the second one reflects the error in continuum placement. 
We did not allow for blending with the HD $(6-0)$ $R(0)$ line at 1031.93 \AA.
($d$) $v_\infty$ from \citet{prinja98}.
($e$) Only data from orbital night were included to reduce airglow contamination.
($f$) $v_\infty$ from \citet{howarth97}.}.
\end{deluxetable}

\clearpage
  {\scriptsize 
  Top: {\fuse} spectra of {\sk{-69}{124}} in the vicinity of 
  {\ion{O}{6}} obtained at two times separated by 62 days.
  The rest wavelength and {\vinf} are indicated for the 
  {\ion{O}{6}} doublet by horizontal bars. The observed differences 
  between $\sim$1027 and 1035 \AA\ are due to wind variability from both components of the doublet.
  The horizontal thick dotted line shows where the Galactic and LMC interstellar components would 
  be expected.
  Bottom: Ratio of the fluxes plotted in the
  rest frame velocity of each component of the doublet. 
  Note that while the wind variability does not affect the interstellar 
  \ion{O}{6} $\lambda1038$ feature, 
  the middle diagram shows clearly how the high velocity
  \ion{O}{6} $\lambda1038$ wind variability can interact with the interstellar 
  \ion{O}{6} $\lambda1032$ feature.
  The dotted line indicates {\vinf}, while the extent of the variation is 
  shown by the solid thick vertical lines. 
  Error bars show the typical $1\sigma$ error in the flux (top) or flux ratio
  (bottom).
  The positions of strong airglow lines due to {\ion{H}{1}} and {\ion{O}{1}} 
  emission are indicated in both panels.
  Note that while the wind variability does not affect the interstellar 
  \ion{O}{6} $\lambda1038$ feature, 
  the middle diagram shows clearly how the high velocity
  \ion{O}{6} $\lambda1038$ wind variability can interact with the interstellar 
  \ion{O}{6} $\lambda1032$ feature.
  The dotted line indicates {\vinf}, while the extent of the variation is 
  shown by the solid thick vertical lines. 
  Error bars show the typical $1\sigma$ error in the flux (top) or flux ratio
  (bottom).
  The positions of strong airglow lines due to {\ion{H}{1}} and {\ion{O}{1}} 
  emission are indicated in both panels.}
\newpage 

\newpage

\end{document}